\documentclass[pra,preprintnumbers,amsmath,amssymb,twocolumn]{revtex4}
\usepackage{amsmath}
\usepackage{graphicx}
\usepackage{amssymb}
\makeatletter

\@ifundefined{textcolor}{}
{
 \definecolor{BLACK}{gray}{0}
 \definecolor{WHITE}{gray}{1}
 \definecolor{RED}{rgb}{1,0,0}
 \definecolor{GREEN}{rgb}{0,1,0}
 \definecolor{BLUE}{rgb}{0,0,1}
 \definecolor{CYAN}{cmyk}{1,0,0,0}
 \definecolor{MAGENTA}{cmyk}{0,1,0,0}
 \definecolor{YELLOW}{cmyk}{0,0,1,0}
 }

\newcommand{\carbon}{$^{13}$C}
\newcommand{\siliconspin}{$^{29}$Si}
\newcommand{\proton}{$^{1}$H}

\newcommand{\tone}{$T_1$}
\newcommand{\ttwo}{$T_2$}
\newcommand{\xenonspin}{$^{129}$Xe}

\begin{document}

\title{In-vivo magnetic resonance imaging of hyperpolarized silicon particles}

\author{M.~C. Cassidy$^{1}$}
\author{H. R. Chan$^{2}$}
\author{B.~D. Ross$^{2}$}
\author{P.~K. Bhattacharya$^{2,3}$}
\author{C.~M. Marcus$^{4,5}$}

\affiliation{$^1$School of Engineering and Applied Sciences, Harvard University, Cambridge, Massachusetts 02138, USA\\
$^2$Huntington Medical Research Institutes, Pasadena, California 91105, USA\\
$^3$Experimental Diagnostic Imaging, The University of Texas MD Anderson Cancer Center, Houston, Texas 77030, USA\\ $^4$ Department of Physics, Harvard University, Cambridge, Massachusetts 02138, USA \\ $^5$ Center for Quantum Devices, Niels Bohr Institute, University of Copenhagen, 2100 Copenhagen, Denmark }

\begin{abstract}

Silicon-based micro and nanoparticles have gained popularity in a wide range of biomedical applications due to their biocompatibility and biodegradability \emph{in-vivo}, as well as a flexible surface chemistry, which allows drug loading, functionalization and targeting. Here we report direct \emph{in-vivo}~imaging of hyperpolarized \siliconspin~nuclei in silicon microparticles by MRI. Natural physical properties of silicon provide surface electronic states for dynamic nuclear polarization (DNP), extremely long depolarization times, insensitivity to the \emph{in-vivo} environment or particle tumbling, and surfaces favorable for functionalization.  Potential applications to gastrointestinal, intravascular, and tumor perfusion imaging at sub-picomolar concentrations are presented. These results demonstrate a new background-free imaging modality applicable to a range of inexpensive, readily available, and biocompatible Si particles.
\end{abstract}

\maketitle

\section*{Introduction}

Nanomedicine is an emerging field that offers great promise in the development of non-invasive strategies for the imaging, diagnosis and treatment of disease. The development of platform technologies that suit a wide range of potential applications is crucial due to the complexity of the \emph{in-vivo} response coupled with time consuming and expensive toxicity studies\cite{Nel,WCWChan,Xia}. Silicon and its oxide derivatives are one such material that has emerged for targeting and drug delivery that may suit a range of intravascular and gastrointestinal applications\cite{Tasciotti,Park, Tanaka, Singh, Mann,Tasciotti2}. To date, \emph{in-vivo}~imaging and tracking of silicon particles has been realized via confinement-enhanced optical activation\cite{Park} or by the incorporation of imaging agents such as fluorescent markers\cite{Tasciotti}, paramagnetic compounds for conventional magnetic resonance imaging (MRI)\cite{Ananta, Gu}, or radionuclides for positron emission tomography (PET)\cite{Kauzlarich_PET}.

Magnetic resonance imaging (MRI) is an attractive technique for both \it{in-vivo}\rm~studies and clinical diagnosis as it is non-invasive, yields high anatomical resolution, and requires no ionizing radiation. However, the small magnetic moment of atomic nuclei means that large numbers of nuclei are required for imaging under conventional conditions. One approach to increasing the MRI signal is to increase polarization far beyond its equilibrium value---a technique known as hyperpolarization. For instance, hyperpolarized noble gases have shown great promise in structural imaging of the lungs\cite{Albert, Fain, RON}, while hyperpolarized $^{13}$C and $^{15}$N metabolites have been used for studies of metabolism and pH in oncology\cite{Brindle, Brindle2, vingeron, pratip}.  A key requirement for hyperpolarization is the transfer of angular momentum from a highly polarized source to the target nuclei. For example, the source of angular momentum in the noble gas experiments is an optically excited vapor of alkali metal\cite{Happer}. For liquid state metabolites, the source is the spin of free radicals (unpaired electrons) combined with the metabolite of interest for low temperature dynamic nuclear polarization (DNP)\cite{griffin, AL, golman}, or parahydrogen in combination with an organo-metallic catalyst\cite{golman, pratip}. 

MRI using silicon is particularly attractive for several reasons. First, the nuclear magnetic moment of \siliconspin~is close to that of $^{13}$C and $^{15}$N, putting it within tuning range of commercial multinuclear MRI systems. Second, direct imaging of \siliconspin~is essentially background free, as the body contains only trace quantities of Si naturally. Third, because native Si consists of a dilute (4.6\%) concentration of spin-1/2 \siliconspin~nuclei in a nuclear-spin-free lattice, the weak dipolar interaction between \siliconspin~nuclei, and their insensitivity to crystalline electric fields (unique to spin 1/2 nuclei) results in remarkably long relaxation times (\tone), up to several hours in high purity samples\cite{Shulman, Aptekar}, coherence times (\ttwo) of several seconds\cite{Ladd05}, and an insensitivity to lattice orientation or rotation. Fourth, while there are few unpaired electrons within bulk of high purity crystalline Si, surface defects comprising unpaired electrons are prevalent at the naturally occurring silicon/silica (silicon dioxide) interface (Fig.~1a). Surface defects are ideal for DNP\cite{Dementyev}, and, because they are remote from most \siliconspin~nuclei, do not lead to rapid relaxation of nuclear hyperpolarization. 

\begin{figure}
\centering\includegraphics[width=2.9 in ]{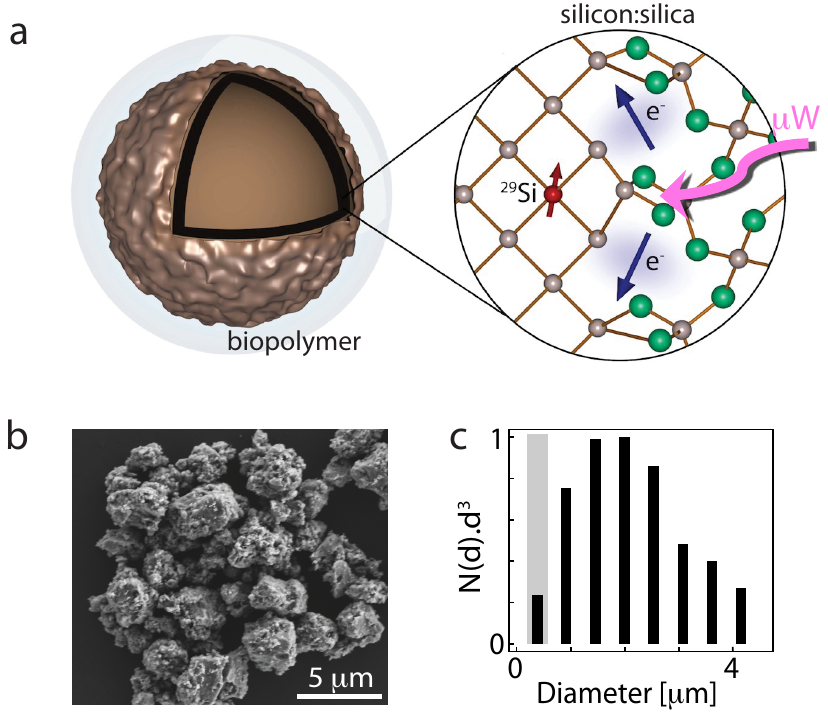}
\caption{\bf{Characterization of silicon particles.}\rm~(a) Schematic diagram showing the compound core-shell structure of the silicon particles surface functionalized with biocompatible polymers. Unbonded electrons naturally occurring at the silicon-silica interface near the particle surface polarize nearby \siliconspin~nuclear spins through DNP. (b) SEM image of the silicon particles used in this experiment. (c) Volume weighted particle size distribution obtained by SEM image analysis showing an average particle size of $2\mu\rm{m}$ (black bars). The average crystallite size is $\sim$ 350 nm (grey bar), as determined by powder XRD analysis.}
\end{figure}

\section*{Materials characterization and nuclear spin dynamics of silicon particles}

In this study, we used commercially obtained high purity silicon particles with a mean diameter of 2~$\mu$m (Fig.~1b-c,  see Supplementary Information S1.1). Each particle consists of a number of crystalline cores ($\sim$~350~nm diameter, determined by powder x-ray diffraction (XRD) analysis, see Supplementary Information S1.2 and Fig.~S1) surrounded by amorphous silicon and a silica shell. Particles were either used in their as-supplied condition or were surface functionalized with 3-aminopropytriethoxysilane (APTES) and polyethylene glycol (PEG) for increased biocompatability\cite{Godin} and hydrostability (see Methods). The complex surface structure of the particles (Fig.~1b) results in a high surface-to-volume ratio (0.9~m$^2$/g, determined by nitrogen adsorption-desorption isotherms, see Supplementary Information S1.3), giving a large density ($\sim~$10$^{13}$ cm$^{-2}$, determined by electron spin resonance, see Supplementary Information S1.4) of defect-bound electrons. 

At low temperatures and high magnetic fields, unpaired electrons at the particle surface are highly spin polarized. By applying microwave irradiation slightly below the electron spin resonance frequency, spin flip-flops are driven between nearby dipolar-coupled electrons, leading to a net transfer of spin polarization from the electrons to \siliconspin~nuclei near the surface. Over time, \siliconspin~nuclei in the particle core become hyperpolarized via nuclear spin diffusion from the surface, a process similar to the conduction of heat. Nuclei in the core region are protected from relaxation, the dominant pathway being slow nuclear spin diffusion back to the surface. Provided that the magnitude of the external field exceeds the dipolar field of the \siliconspin~nuclei, the time scale for this process is independent of ambient magnetic field\cite{MYL} and temperature\cite{AbragamBook}, depending only on hyperpolarization time ($\tau_{pol}$), particle size and the purity of the particle\cite{Aptekar}. This allows the hyperpolarized particles to be transported through an environment of varying magnetic fields from the polarizer to the MRI system over several minutes without a significant loss of polarization. This is in contrast to hyperpolarized materials in the liquid state such as  $^{13}$C~labelled metabolites, where molecular and bond rotations occurring at a frequency equal to the nuclear Lamor frequency induce additional nuclear spin-lattice relaxation\cite{AbragamBook}. 

We used a home-built DNP polarizer operating at a magnetic field of 2.9~T and temperature of 3.4~K located adjacent to a 4.7~T animal MRI system (see Methods and Supplementary Information S2.1 and Fig.~S2). A comparison of \siliconspin~spectra for dry particles and \it{in-vivo}\rm~measurements in the MRI system after a depolarization time $\tau_{dep}$ =  30~min following hyperpolarization is shown in Fig.~2a. Both spectra show a line width (full width at half maximum) of $\sim$~400~Hz, consistent with the dipolar line width for crystalline silicon. A slight additional broadening is seen in the dry sample, presumably corresponding to \siliconspin~nuclei close to the particle surface. We attribute the absence of this additional broadening in the \it{in-vivo}\rm~case to nearby \proton~nuclear spins in the \it{in-vivo}\rm~ environment that cause surface \siliconspin~nuclei to relax on times scales of roughly \tone~of the bordering \proton~nuclei. Figure~2b shows the decay of \siliconspin~polarization for dry and \it{in-vivo}\rm~particles using a variable flip angle sequence following hyperpolarization for a time $\tau_{pol}$ = 18 h (see Methods). Despite the difference in absolute \siliconspin~signal between the two measurements, the characteristic depolarization time constant $T_{1,dep}$ is approximately the same for the dry ($T_{1,dep} = 38 \pm 2$ min) and \it{in-vivo}\rm~cases ($T_{1,dep} = 39 \pm 3$ min). The difference in absolute signal is attributed to sample loss in the dilution and injection procedures, a reduced concentration of silicon particles within the active region of the coil, as well as a contribution from enhanced surface relaxation. By comparing the signal strength to that of a co-located \siliconspin~phantom (see Methods) we estimate a \siliconspin~nuclear polarization of roughly 1\% in the imager. 

\begin{figure}
\centering\includegraphics[width=3 in]{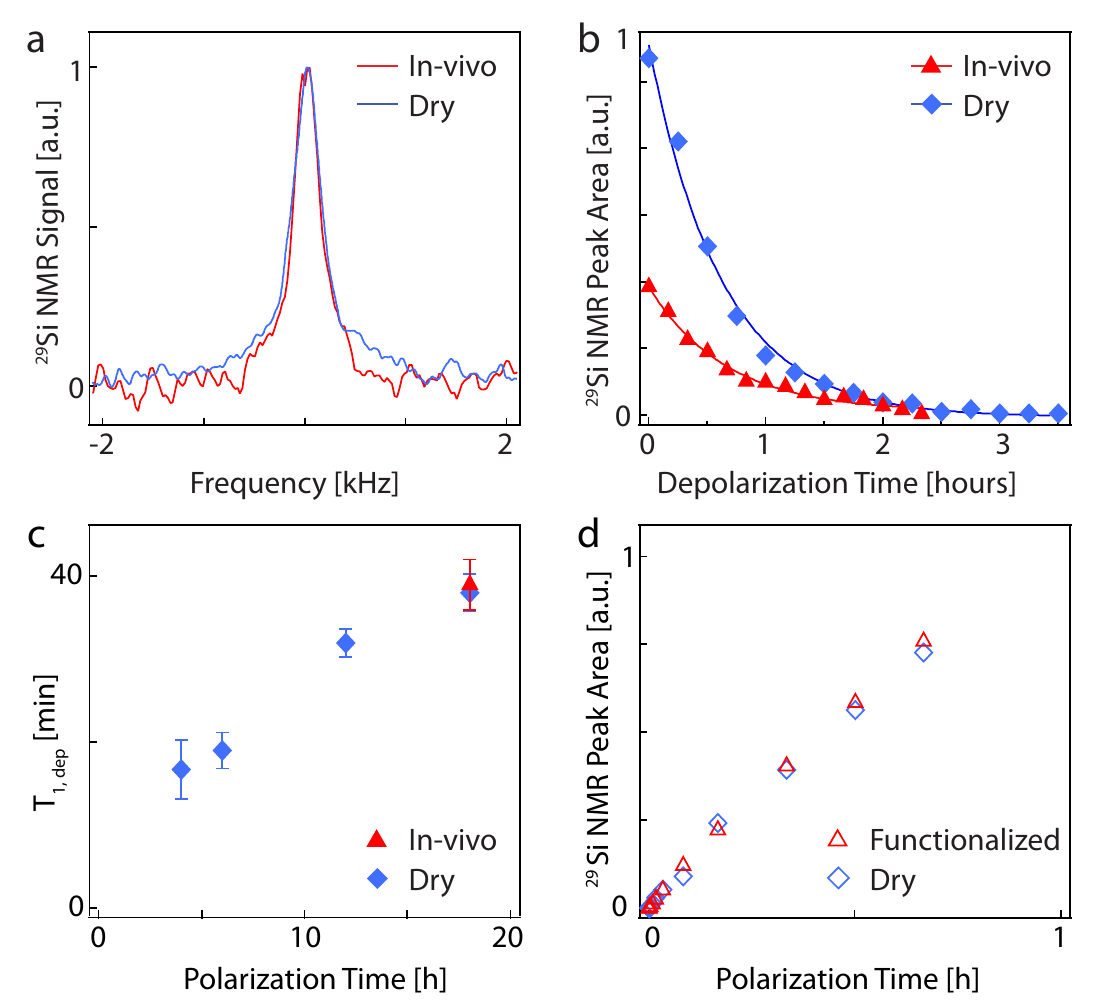}
\caption{\bf{\siliconspin~nuclear spin dynamics in silicon particles.}\rm~(a) \siliconspin~ NMR spectrum of hyperpolarized particles recorded \it{in-vivo}\rm~(red) and dry (blue) at 30 min delay.  (b) Decay of \siliconspin~nuclear polarization in hyperpolarized silicon particles recorded \it{in-vivo}\rm~ (red) and dry (blue). Both show a characteristic decay $T_{1,dep}$ $\sim40$ min for hyperpolarization time $\tau_{pol}$ = 18 h using a variable flip angle acquisition. The difference in absolute signal is mainly due to reduced concentration within the active region of the coil. (c) Characteristic \siliconspin~depolarization time constants ($T_{1,dep}$) in dry particle phantoms recorded in the imager for polarization times of 4, 6, 12, and 18 hours. (d) Time evolution of the \siliconspin~nuclear polarization under DNP conditions for particles with and without surface functionalization with PEG. The signal is normalized by sample weight.}
\end{figure}

The observed depolarization time constants are significantly longer than the ensemble-averaged characteristic spin-lattice relaxation time constant, \tone~$= 29 \pm 5$ min, of the sample measured by a saturation recovery method  at 2.9~T  and 300~K (see Supplementary Information S2.2 and Fig.~S3). Isolated electronic defects within the particle core are ineffective for DNP and act only as local relaxation sites. Once the temperature is raised and magnetic fields is lowered for transfer to the imager, these internal defects and the surface defects used for DNP all act as strong points of relaxation on local \siliconspin~nuclei. Depolarization measurements following DNP are therefore dominated by contributions from nuclei far from local relaxation points, giving a longer characteristic time constants for depolarization from the hyperpolarized state than for polarization under Boltzmann conditions. The depolarization time constants are lengthened with increasing hyperpolarization time ($\tau_{pol}$) (Fig.~2c), and could be further improved by increasing the polarization efficiency or improving the internal crystal structure of the particles (see Supplementary Information S2.3 and Fig.~S4). No significant difference is observed in the \siliconspin~hyperpolarization rate for short time periods (Fig.~2d), where the hyperpolarization may be affected by nearby \proton~nuclei within the functionalizing moieties. This is not unexpected, as the APTES linker contains few spinful nuclei and provides a buffer between the particle surface and the \proton~rich PEG.

\section*{In vitro and \it{\bf{in-vivo}}\rm\bf~imaging}

To establish the practical time window for obtaining hyperpolarized MRI images, we imaged a millimeter-scale phantom constructed from segments of dry silicon particles separated by teflon spacers (Fig.~3a). Images were recorded at delay times $\tau_{dep}$ = 0, 30, 60, and 90 min after placing the phantom in the MRI imager, using separate hyperpolarizations for each time interval. The phantom image is clearly visible at $\tau_{dep}$ = 0, 30 and 60 min for a single average with a signal to noise ratio (SNR) greater than 12. After 90 min, only the top three segments are visible (SNR > 3),  probably due to a variation in polarization across the sample resulting from an uneven microwave field distribution in the polarizer.

\begin{figure}
\centering\includegraphics[width=3 in]{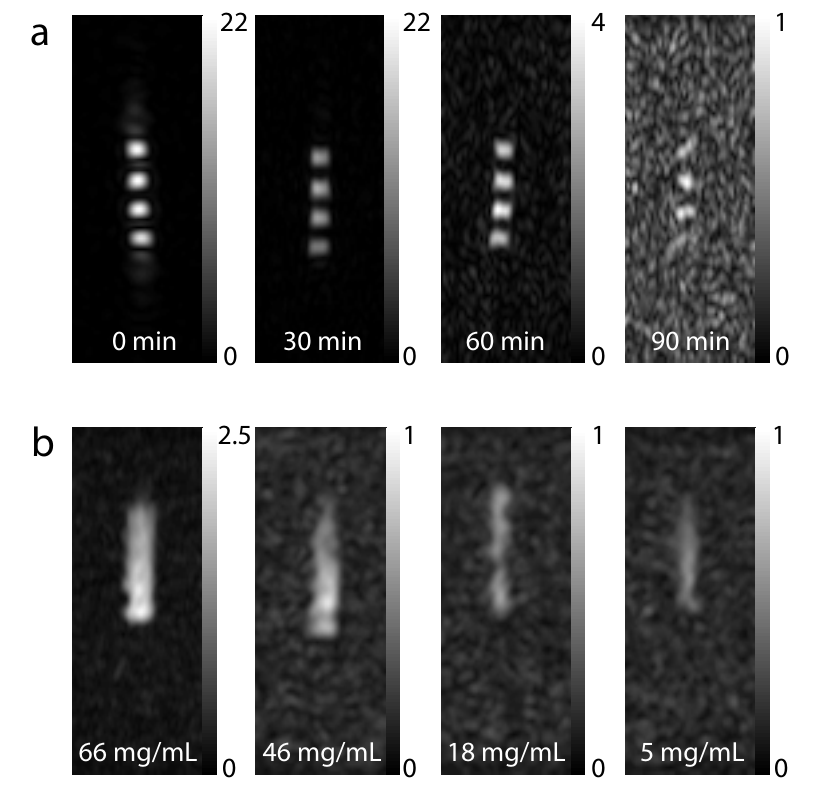}
\caption{\bf{\siliconspin~Phantom Imaging.}\rm~(a) \siliconspin~MRI of a segmented phantom containing silicon particles ($\sim~$15~mg in each segment) with a delay $\tau_{dep}$ of 0, 30, 60 and 90 min from when the phantom was loaded into the magnet until imaging. (b) \siliconspin~MRI  of a cylindrical concentration phantom with particle concentrations of 66~mg/mL, 46~mg/mL, 18~mg/mL and 5~mg/mL. Images have been cropped from 40~mm to 15~mm in the horizontal dimension after processing.
}
\end{figure}

Figure~3b shows the concentration sensitivity for the hyperpolarized particles. The images are of 250~$\mu$L cylindrical phantoms containing between 66 and 5~mg/mL of silicon particles dispersed in ethanol. A sensitivity limit is set by the silicon particle concentration in the outer two columns of the least concentrated sample. Assuming a uniform distribution of particles within the phantom and taking into account the cylindrical projection onto the plane, this corresponds to a visible single pixel concentration of $\sim$10~$\mu$g of hyperpolarized silicon particles per pixel. For a particle diameter of  2~$\mu$m, we estimate an absolute sensitivity of $\sim5~\rm{x}~10^5$ individual particles/pixel, or a particle molar sensitivity of $\sim4.2~\rm{x}~10^{-13}$ mol/L at a depth of 1 cm from the face of the $\sim$4~cm imaging coil. This corresponds to an absolute \siliconspin~sensitivity of $\sim7~\rm{x}~10^{-3}$ mol/L, which is comparable in sensitivity with other reported paramagnetic MRI nanoparticle agents\cite{bouchard} and significantly better than $\rm^{19}F$ labelled cells\cite{Ahrens, Bonetto}.  
  
\begin{figure*}[t!]
\centering\includegraphics[width=\textwidth]{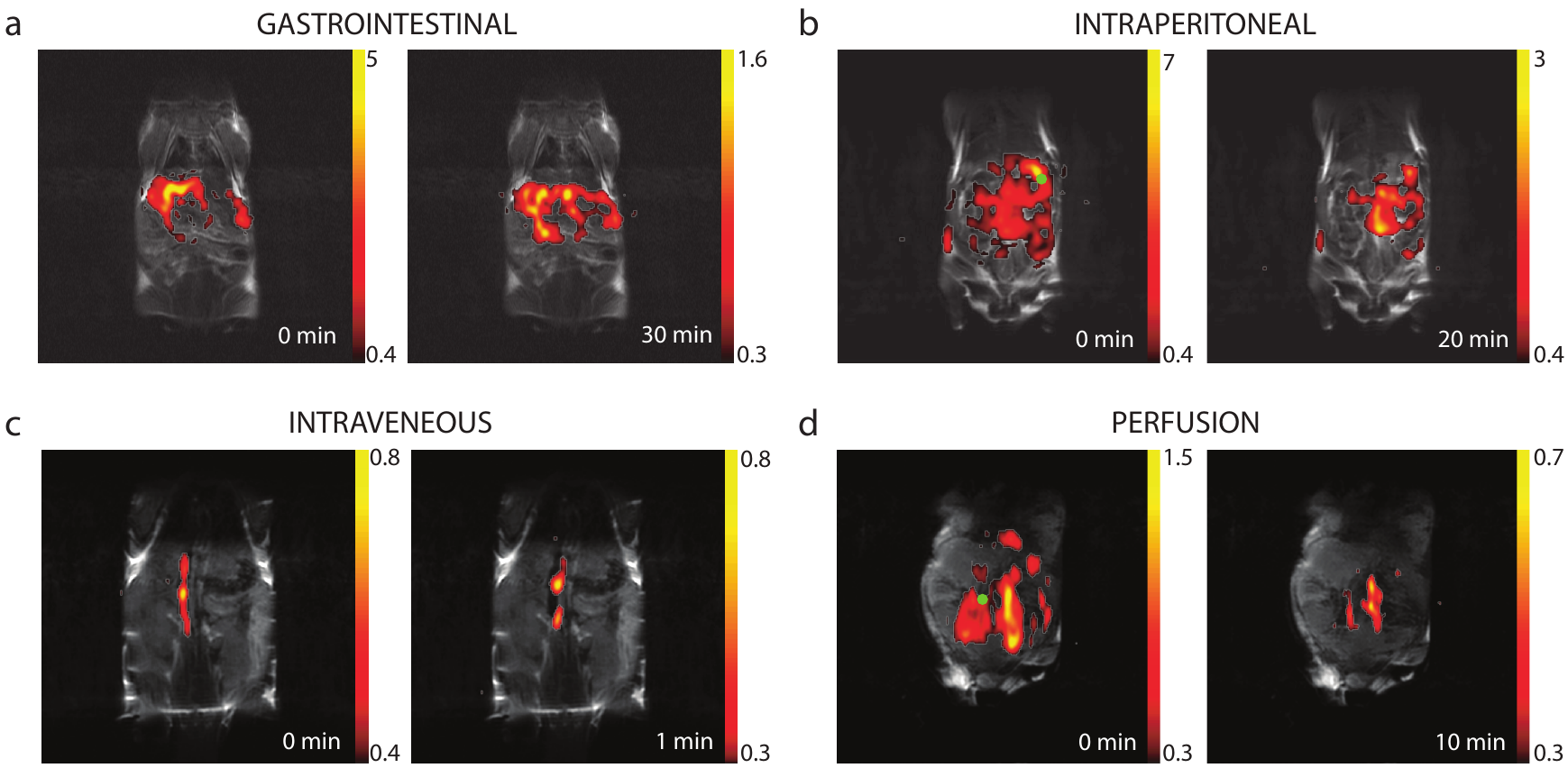}
\caption{\bf{\emph{In-vivo} applications of \siliconspin~MRI using hyperpolarized silicon particles.}\rm~\emph{In-vivo} coregistered \proton: \siliconspin~MRI of hyperpolarized silicon particles administered (a) intragastrially, (b) intraperitoneally, and (c) intravenously via a tail vein catheter. (d) Demonstration of perfusion imaging using \siliconspin~MRI of hyperpolarized silicon particles injected into the tumor of a prostate cancer (TRAMP) animal. In (b) and (d) the the green circle shows the approximate location of the catheter. Later stage images are shown in the Supplementary Information.}
\end{figure*}

We have carried out a number of \it{in-vivo}\rm~imaging studies to demonstrate the versatility of hyperpolarized \siliconspin~MRI. Figure 4a shows coregistered \proton:\siliconspin~images of hyperpolarized silicon particles injected via a catheter into the gastrointestinal (GI) tract of a mouse. Immediately following the injection (< 1 min), the image shows the particles accumulating in the stomach and the duodenum. After 30 min, the \siliconspin~image shows the particles have moved further into the small intestine of the animal, further elucidating internal structure. Delivery into the intraperitoneal (IP) cavity reveals detailed external structure of the GI tract imaged over 40 min (Fig. 4b, also see Fig. S6). 

Diagnostic imaging of the small intestine is notoriously difficult as it is inaccessible to traditional endoscopic techniques. Current non-invasive GI imaging techniques using computerized tomography (CT) or \proton~MRI, with a combination of positive and negative contrast agents delivered intravenously and intragastrically for delineating internal structure\cite{Radiology}. For MRI, the complex air-tissue interfaces of the GI region cause susceptibility gradients that dominate over the change in signal induced by the paramagnetic agent. Silicon has been studied as a nutritional supplement\cite{canham, macdonald} and is a common food additive (up to concentrations of 2\%) in its oxide form\cite{FDAregs}. Hybrid gold-silicon nanoparticles are currently under investigation as targeted colorectal optical imaging agents and show no evidence of toxicity\cite{thakor}. Hyperpolarized silicon particles appear well suited for structural and targeted imaging of GI disease by MRI.

For particles administered intravenously (IV) via a tail vein catheter (Fig.~4c, also see Fig.~S7), the vena cava is visible immediately after injection and continues to be visible one and two minutes later. The PEG chain used to functionalize the particles was chosen as it is known to enhance the circulation of similarly sized particles\cite{Godin}. The high concentration of particles carried in the vena cava, and the proximity of the vena cava to the imaging coil results in this signal dominating over signals from other areas in the animal. After two minutes, some particle accumulation is observed in the organs of the mononuclear phagocyte system (MPS), consistent with accumulation patterns of similar sized materials imaged optically\cite{Park}. 

As a preliminary demonstration of perfusion imaging with hyperpolarized silicon MRI, we administered hyperpolarized silicon particles into the prostate tumor of a TRansgenic Adenocarcinoma of Mouse Prostate (TRAMP) mouse (Fig.~4d, also see Fig.~S8). The particles used in this study are too large to enter the cellular structure of the tumor, however they accurately map the blood vessel microstructure inside the tumor, as well as the direction of blood flow within the tumor. Initially the \siliconspin~MRI shows accumulation at the lower right quadrant of the tumor while later images show a higher concentration in the upper right quadrant. There was little \siliconspin~signal observed on the left side of the tumor at any imaging time, suggesting an area of necrosis. This observation was confirmed visually post-mortem.

A key advantage of hyperpolarized \siliconspin~MRI is biocompatability. Silica is frequently used as a biocompatible coating for other nanoscale probes\cite{coating, coating2}, and several studies have reported negligible toxicity of silicon particles\cite{Park, thakor}. In these investigations, the clearance mechanism has been attributed to degradation of silicon and silica into soluble silicic acid followed by excretion within a few days. Our preliminary study of in-vivo toxicity of these particles up to the imaging concentrations show no decrease in animal viability over a two week period for particles administered via GI, IP or IV routes. Comparison of MPS organ tissue sections between those harvested three hours after IV administration and two weeks (Fig.~S9) show that these silicon particles appear to be efficiently cleared within the two week period.

\section*{Conclusion}

Several orders of magnitude improvement in the concentration threshold and imaging time-frame presented here are possible. Eliminating isolated defects in the internal crystal structure and chemically enhancing the surface defect density, would result in more efficient hyperpolarization and significantly longer depolarization time constants. The hyperpolarization mechanism via surface defects is applicable to silicon particles of any size, with the time window available for imaging ultimately set by the hyperpolarization time and the distance the nuclear spin hyperpolarization has to diffuse to relax. Previous studies indicate that the depolarization times observed here ($\sim$ 40 min) do not decrease significantly as particle size is reduced by an order of magnitude below those used in this study\cite{Aptekar}, and we have recently engineered 10~nm particles with \tone~times greater than 10 minutes\cite{Atkins}.  Isotopic enrichment, though not yet commercially available, reduces the hyperpolarization and depolarization time constants by a theoretical factor of $n^{2/3}$, where $n$ is the isotopic fraction, but brings about a linear increase in signal. Although, we have studied enriched \siliconspin~samples (92\% \siliconspin) in other studies \cite{Cassidythesis} and observe that the decreases in hyperpolarization and depolarization times qualitatively agree with theory.  

Because the time scale for depolarization does not depend on surface functionalization, targeting strategies developed for silica coated nanovectors can be applied to particles optimized for hyperpolarized \siliconspin~MRI. Surface doping may allow for extremely large nuclear polarizations to be generated on fast timescales\cite{McCamey, Bagraev} through combined optical-microwave techniques. Additionally, as the \it{in-vivo}\rm~$T_{1,dep}$~times are at least two order of magnitude longer than those of hyperpolarized \carbon\cite{golman} and \xenonspin\cite{xenont1}, it may be possible to use the nuclear hyperpolarization within the particles to store polarization (as a battery), then transfer polarization to read-out nuclei with higher gyromagnetic ratio via established cross-polarization techniques. 

Several other approaches to background-free imaging are currently under development. Most notable is $\rm^{19}F$, which has been widely investigated as a contrast agent both in drugs and in perfluorocarbon (PFC) nanoparticles\cite{Lanza}. Its 100\% natural abundance and high gyromagnetic ratio together with a lack of \it{in-vivo}\rm~background signal allows it to be imaged without hyperpolarization. However \it{in-vivo}\rm~studies using $\rm^{19}F$ typically require large numbers of averages\cite{Bonetto}, compared to the singles-shot images shown here. Such long averaging times do not allow for real-time tracking of particles \it{in-vivo}\rm, and are suited best to accumulation studies over long time periods where the particle location has reached a steady state. Additionally, unlike Boltzmann polarized materials such as $\rm^{19}F$, where the signal is dependent on strength of the magnetic field used for imaging, the hyperpolarized \siliconspin~nuclei can be easily imaged with high sensitivity at low magnetic fields using affordable permanent magnet or inductive coil MRI systems.   

A key concern towards clinical applications of hyperpolarized \siliconspin~imaging may be the specific absorption rate (SAR) issues associated with the use of fast spin-echo imaging techniques used in this work. We note that although spin-echo imaging gives a significant increase in signal, we have also employed gradient-echo imaging techniques for \siliconspin~MRI\cite{Aptekar}. As the power deposited scales quadratically with the Lamor frequency of the nucleus being imaged, the low gyromagnetic ratio of silicon together with the possibility of low-field imaging alleviates many of these concerns. Additionally, research into fast MRI techniques such as restricted k-space acquisition and compressed sensing\cite{Lustig}, as well as hardware developments that minimize electric fields generated in imaging coils are extremely applicable to \siliconspin~MRI.

In summary, we have demonstrated \it{in-vivo}\rm~imaging of hyperpolarized silicon particles by MRI, and investigated a range of potential applications. By choosing particles with a crystalline Si core protected a surface oxide, the polarization generated in the particles is unaffected by the \it{in-vivo}\rm~environment or particle tumbling. Magnetic resonance offers not only the possibility of structural imaging, but also also functional imaging of flow, particle binding, or the local environment through surface functionalization. The lack of \it{in-vivo}\rm~\siliconspin~background signal has the potential for quantification of the \siliconspin~magnetic resonance label, opening up the possibility to track the particle biodistribution systematically, analogous to the use of nuclear tracers.

\section*{Methods}

\subsection{Preparation of silicon particles}
Silicon powder ($2~\mu \rm{m}$ APS, 99.9985\% elemental purity, Alfa Aesar) was used both as supplied and following surface functionalized with polyethyleneglycol (PEG) for biocompatibility.
Particles were suspended in acidified ethanol (pH 3.5, Sigma 100\%) and placed in an ultrasonic bath for 5 min. 3-aminopropyltriethoxysilane (APTES, Sigma 99\%) was added at a concentration of 100~$\mu \rm{L}$/g of silicon and the sample shaken on a plate shaker for 24~h. Excess silanes were removed by centrifiguation and rinsing, and the particles resuspended in ethanol buffer. NHS-dPEG4-(m-PEG12)3-ester (Quanta Biodesign) was added at a concentration of 100~mg/g of silicon and the sample was shaken at 45$^\circ$~C for 12 h. The particles were then rinsed in acidified ethanol buffer and concentrated by centrifugation. Samples ($\sim$100 mg) were packed into thin wall Teflon tubes that withstood the rapid thermal cycling involved with the polarization process, before being loaded into the polarizer.

\subsection{\siliconspin~dynamic nuclear polarization}
DNP was performed at 3.4 K in a custom built polarizer operating at 2.9 T for hyperpolarization times $\tau_{pol}$ in the range 4 - 24 h. Frequency modulated microwave irradiation was applied between 80.83 - 80.91~GHz (10 kHz ramp modulation) with a 2 W microwave source (Quinstar) located at room temperature and coupled to the sample via a mm waveguide. Details of the polarizer construction and frequency modulation technique are described elsewhere\cite{Cassidythesis}.

\subsection{Sample transfer and dissolution}
Solid samples were removed from the polarizer and transported to the face of the magnet for dissolution. Phosphate buffered saline (0.5 - 1 mL) or ethanol was added and the sample suspended by manual agitation. The total transport and dissolution time was approximately 2 min.

\subsection{NMR Experiments}

The characteristic depolarization time constant $T_{1,dep}$ was measured in the animal scanner using a variable flip angle sequence that preserved the intrinsic \tone~decay of the sample\cite{Zhao}. For $N$ experiments, the flip angle for the $n^{th}$ pulse was calculated by $\theta_n = tan^{-1}(\frac{1}{N-n})$.

All \siliconspin~spectra were exported to Igor Pro (v 6.22, Wavemetrics) for analysis. A 5~ms exponential filter was applied to the complex time domain data before Fourier transform. 

\subsection{MRI Experiments}
All phantom and animal MRI experiments were performed at 4.7 T in a horizontal bore animal scanner (Bruker Biospin) outfitted with a high-resolution gradient set. Control of the experiment was performed using the preclinical software Paravision (v 3.0.2 Bruker Biospin). 
A dual coil setup was used for co-registered \proton~:\siliconspin~imaging. A \proton~volume coil was used for sample placement and anatomical imaging. \siliconspin~imaging and spectroscopy was performed with a custom built surface coil (38 mm ID) that was placed on the abdominal region of the animal. The coil size was slightly increased (45 mm ID) for the TRAMP animal due to the size of the tumor. A small ($\sim1$~mL) phantom of tri-methyl silanol (TMS, 99\% Sigma) was located in the volume coil at a depth 1~cm from the face of the surface coil as a \siliconspin~spectroscopic reference. For phantom imaging, the phantom was placed in a holder centered in the volume coil with a vertical displacement of 1~cm above the face of the surface coil. 
 In Fig. 2(a), the phantom size was $\sim$~2 x 14 pixels within the 64 x 64 imaging matrix (individual segments were $\sim$~2 x 2 pixels). For Fig. 2(b) the phantom size was $\sim$~6 x 22 pixels within the 64 x 64 imaging matrix.
 
Anatomical imaging was performed with a standard \proton~multiple-slice multiple-echo (MSME) spin echo sequence. A custom fast spin-echo pulse sequence was written for \siliconspin~imaging to take advantage of the particular decoherence properties of solid silicon. Details of the imaging sequences, as well as the image processing used is described in the Supplementary Information.

\subsection{Mouse handling}
All animal work was performed in accordance with the institutional animal protocol guidelines in place at the Huntington Medical Research Institute, and was reviewed and approved by the Institute's Animal Research Committee. Male BALB/c mice ($\sim$~20 g) (Harlan Laboratory, Livermore CA) were used for the normal experiments. Male TRansgenic Adenocarcinoma of Mouse Prostate (TRAMP) mice (Jackson Laboratory, Bar Harbor, ME) were used for prostate tumor experiments. The animals were anesthetized by face mask (1\% isoflurane in 0.8 L/min oxygen) and a catheter was positioned intragastrically (GI), intraperitoneally (IP) or intraveneously (IV) or within the tumor. The animal was placed into a custom heated holder compatible with the MRI coils and kept anesthetized with a nose cone during the experiment. 

The hyperpolarized silicon particles were suspended in phosphate buffered saline (Baxter Healthcare) before being injected through the catheter with a 25 Gauge syringe. The suspension amount was 0.3 mL for GI and IP, and 1mL for IV. Additionally, unfunctionalized particles used for the GI experiments were first suspended in 0.2mL of ethanol for enhanced surface stabilization. The injection was followed by 0.2 mL of saline to clear the catheter. After the experiment, the animals were sacrificed while still anesthetized and internal organs (spleen, liver, kidneys) were harvested for further study. 

\section*{Acknowledgements}
 This research is supported by the NIH (1R21 CA118509, 1R01NS048589), the NCI (5R01CA122513), the Tobacco Related Disease Research Program (16KT-0044), NSF BISH (CBET-0933015), and the NSF through the Harvard NSEC. M.C.C. acknowledges financial support from the R.G. Menzies Foundation. C.M.M. acknowledges support from the Danish National Research Foundation and the Villum Foundation. We thank L. W. Jones for the gift of the TRAMP animals, N. Zacharias for assistance with histology, M. Lee, L. Robertson and B. D. Armstrong for assistance with the initial stages of the experiment and L. W. Jones, C. Ramanathan and R. W. Mair for many helpful discussions.

\pagebreak
\clearpage

\section*{Supplementary Information}

\renewcommand{\thesection}{S~\arabic{section}}

\section{Materials preparation and characterization}

\renewcommand{\thesubsubsection}{S \arabic{section}.\arabic{subsubsection}}
\setcounter{subsubsection}{0}

\renewcommand{\thefigure}{S\arabic{figure}}
\setcounter{figure}{0}
\setcounter{table}{0}

\subsubsection{Particle size analysis}  The size distributions of the particles were determined by scannning electron microscopy. Dilute suspensions of silicon particles in methanol were sonicated for 10 mins before being pipetted onto a graphite substrate which was mounted on a standard specimen holder with conducting carbon tape. Particle diameters were extracted from analysis of the SEM images with particle-measuring software (Gatan Digital Micrograph). Approximately 500 particles were analyzed, sourced from $\sim~$10 images. From this the volume weighted size distribution $n(d).d^3$ was plotted, showing an average volume weighted particle diameter of $\sim2~\mu$m. 

\subsubsection{Surface area analysis} The particle surface area was determined by nitrogen adsorption-desorption volumetric isotherm (BET) analysis using a Micromeritics ASAP 2020 System.

\subsubsection{Crystallite size measurements} Powder x-ray diffraction measurements (Scintag XDS2000) were performed to analyze the particle crystallinity and average crystallite size. The peaks at $28^{\circ}$, $47^{\circ}$, $56^{\circ}$ and $69^{\circ}$ (Fig.~S1) agree with the peak locations of a known silicon standard ($10~\mu\rm{m}$ APS, NIST). By comparing the broadening of these peaks compared to the standard we estimate that the particles have an average crystallite size of approximately  350~nm.

\begin{figure}
\centering\includegraphics[width=3 in]{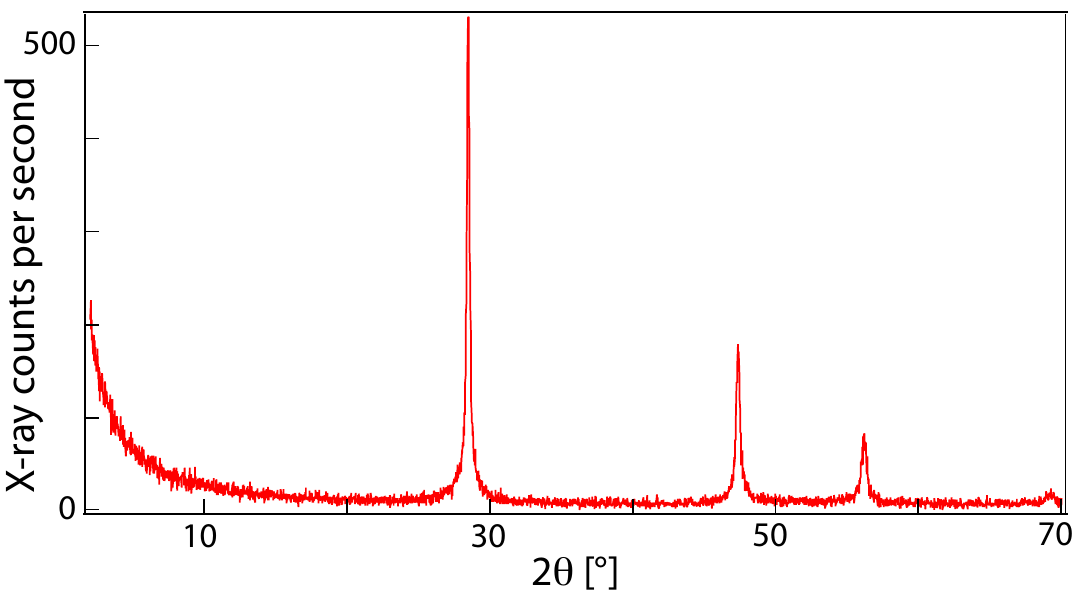}

\caption{\bf{X-ray diffraction measurements}\rm~X-ray diffraction powder spectrum of the silicon particles. The peaks at $28^{\circ}$, $47^{\circ}$, $56^{\circ}$ and $69^{\circ}$ agree with the peak locations of a known silicon standard ($10\mu\rm{m}$ APS, NIST). By comparing the broadening of these peaks compared to the standard we estimate that the particles are approximately 80\% crystalline with an average crystallite size of 350~nm. 
}
\end{figure}

\subsubsection{Defect density measurements} Continuous wave electron spin resonance (ESR) measurements were performed on bulk samples of the particles at 3.4 K (Bruker ElexSys E500). An estimate of the electron spin concentration on the particle surface was made by taking ESR spectra with a piece of phosphorus-doped silicon wafer also inserted in the spectrometer cavity. The addition of the wafer piece resulted in the appearance of additional ESR signal due to the n-type doping. Comparing the peak areas and using the known doping level of the wafer piece ($0.008-0.01 ~\Omega.$ cm) yielded an order-of-magnitude estimate of the mean volume density of defect spins $10^{18}~\rm{cm}^{-3}$. Assuming a surface-to-volume ratio of 2~$\mu$m diameter spheres, the mean volume density corresponds to a surface defect concentration of $10^{13}~\rm{cm}^{-2}$. 

\section{Nuclear spin measurements and modelling}
\setcounter{subsubsection}{0}

\subsubsection{Low temperature DNP measurements}  The \siliconspin~nuclear polarization was measured under dynamic nuclear polarization conditions at 2.9 T in the polarizer using a saturation recovery sequence. A saddle coil that could be tuned across a wide range of temperatures was used for NMR detection. A train of 32 hard $\pi$/2~pulses saturated residual magnetization, and the sample was allowed to hyperpolarize for a time $\tau_{pol}$ before the magnetization was measured with a single $\pi$/2~pulse. The area under the Fourier transformed peak was used as a measure of the nuclear polarization. By comparing the peak area to that acquired at room temperature under Ernst angle conditions, we estimate the \siliconspin~nuclear polarization to be approximately 10~\%~after a hyperpolarization time of $\tau_{pol}$ = 18 h.

 \begin{figure}
\centering\includegraphics[width = 3 in]{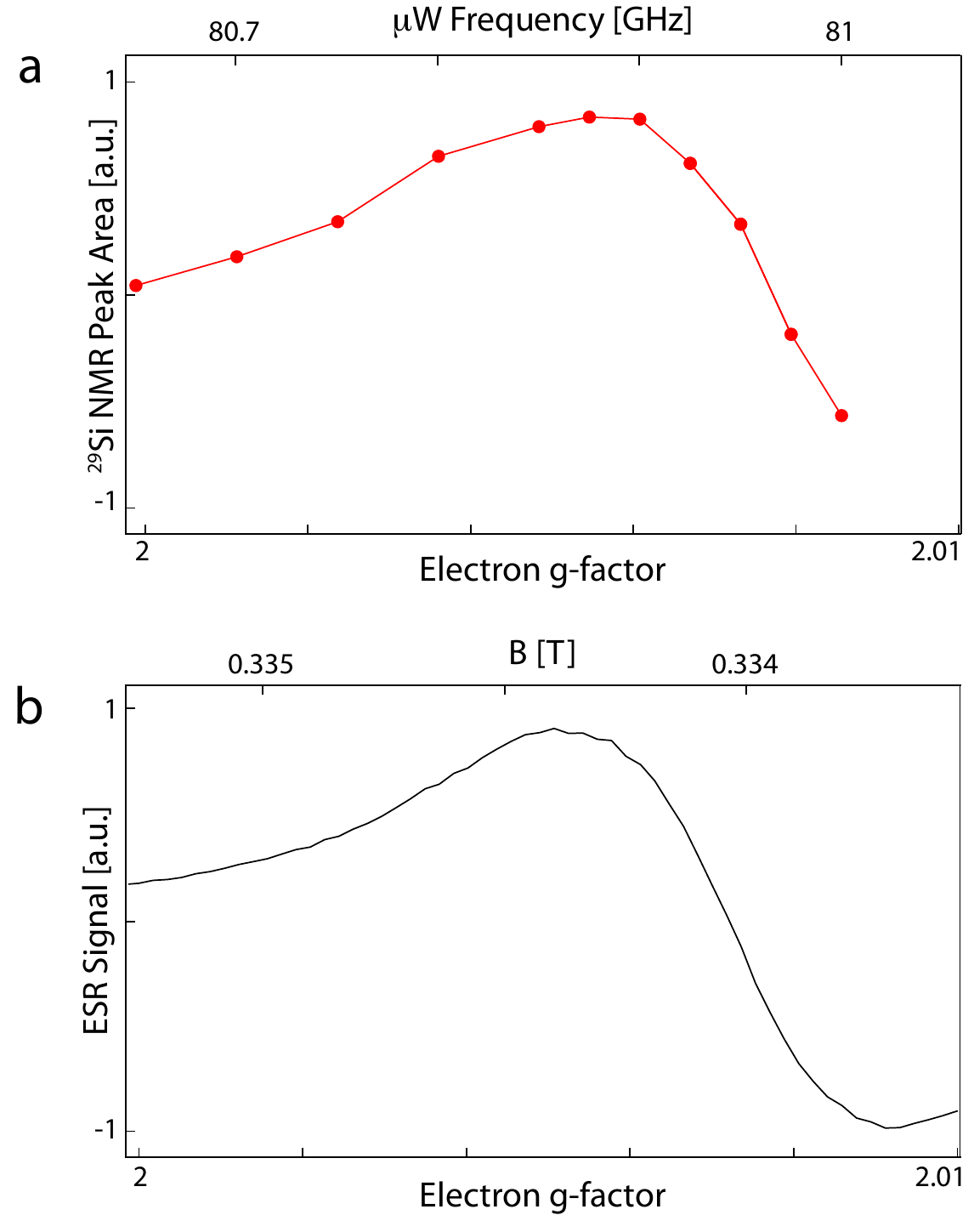}

\caption{\bf{Frequency dependence of dynamic nuclear polarization.}\rm~(a) Microwave frequency dependence of the \siliconspin~nuclear polarization acquired at constant magnetic field at 3.8~K. (b) Electron spin resonance signal as a function of magnetic field acquired at constant microwave frequency at 3.8~K. Both sets of data are plotted over the same range of g-factors. Good agreement is seen between the ESR and DNP despite the difference in magnetic field and microwave frequency for the measurements.
}
\end{figure}       

The microwave frequency dependence of \siliconspin~dynamic nuclear polarization measured at 3.4~K is shown in Fig. S2(a). The ESR signal measured at 9 GHz and 3.4~K is shown in Fig. S2(b) and is plotted over the same range of g-factors as the data in Fig. S4(a).  The \siliconspin~nuclear polarization scales proportional to the derivative of the ESR line with the cross over frequency between negative and positive nuclear polarization occurring at the average value of the electron g-factor of the system (g=2.0065). This is consistent with a DNP mechanism driven by dipolar interactions between a dense bath of electron spins\cite{Khut}. 

For all polarization experiments, frequency modulated microwave irradiation was applied across the peak values in Fig.~S2(b) between 80.83 - 80.91~GHz with a 10 kHz ramp modulation.

\subsubsection{Room temperature \tone~measurements}  The room temperature \tone~of the sample was measured with a saturation recovery Carr-Purcell-Meiboom-Gill (CPMG) sequence\cite{Carr,Meiboom} (Fig.~S3). A train of 32 hard $\pi$/2 pulses saturated residual magnetization, and the sample was allowed to polarize for a time $\tau$. The magnetization was rotated to the transverse plane with a single $\pi$/2~pulse and (N=200) $\pi$~pulses performed. The echo following each of the $\pi$~pulses were Fourier transformed and the height of the peak was used as a measure of the nuclear polarization. The data is fitted with a biexponenetial function $P =P_{0}+\left((1-\alpha)e^{-\tau/T_{1,\rm f}}+\alpha e^{-\tau/T_{1,\rm s}}\right)$ and two time components are extracted. $T_{1,\rm f} = 6 \pm 1$~\rm min. corresponds to nuclei within the shell of the particle, and $T_{1,\rm s} = 29 \pm 6$~min corresponds to nuclei within the core of the particles.

            \begin{figure}
\centering\includegraphics[width=3 in] {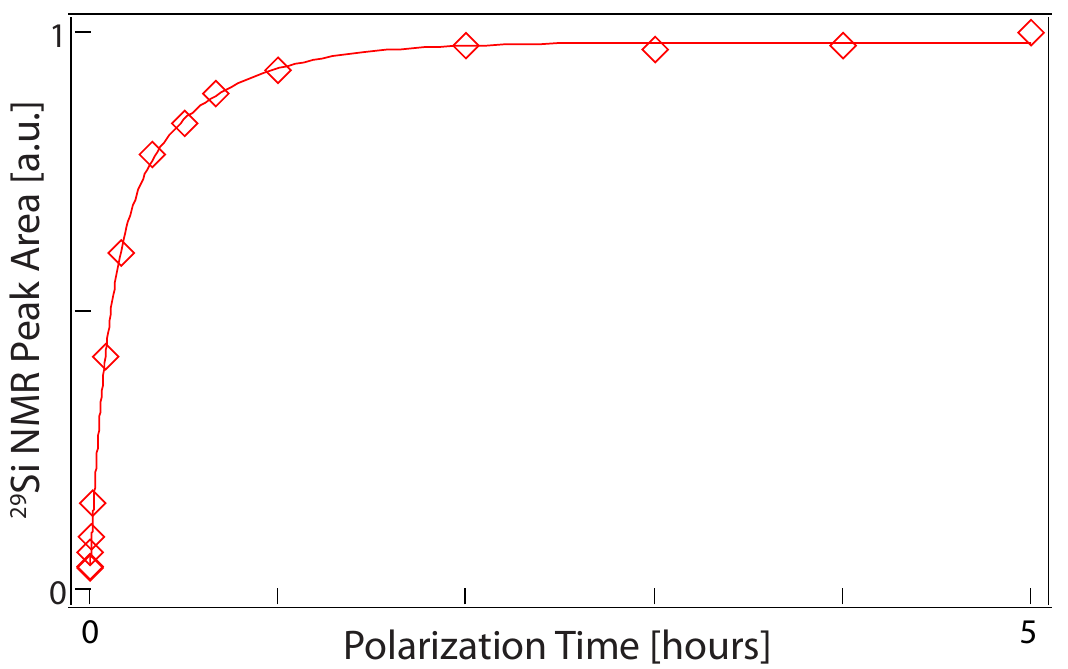}

\caption{\bf{Room temperature \siliconspin~NMR measurements.}\rm~Room temperature \siliconspin~nuclear \tone~measurement of the silicon particles at 2.9 T measured with a saturation recovery CPMG sequence.}
\end{figure} 

\subsubsection{Simulation of depolarization as a function of particle size} We numerically modelled nuclear spin relaxation from a pre-hyperpolarized state for particles with diameters ranging from 50~nm to 2~$\mu$m (Fig. S4). The model assumed that the particles were spherical with natural abundance (4.7\%) \siliconspin~concentration and paramagnetic defects located on the surface. The initial polarization $P(r, t = 0)$ was set to 1~\% uniformly throughout the particle, and the polarization decay simulated as it returned to the Boltzmann polarization at a depolarizing field $B_{\rm dep}$ = 4.7~T and T = 300~K. We considered a volume weighted 1D radial model consisting of 100 distinct spin packets with the decay of polarization in each packet at radial position $r$ described by contributions from nuclear spin diffusion and direct relaxation,

\begin{align}  \frac{dP(r,t)}{dt}= \left ( \frac{dP(r,t)}{dt}\right )_{diff.} + \left (\frac{dP(r,t)}{dt}\right )_{direct}. \end{align} 

The rate of change in nuclear polarization for a spatially inhomogeneous distribution of nuclear polarization is given by, 

\begin{align} \left ( \frac{dP(r,t)}{dt}\right )_{diff.}   = \sum\limits_{\alpha, \beta = 1}^{3} D^{\alpha\beta} ( \frac{\partial^2}{\partial x^\alpha \partial x^\beta} ) P(r,t),\end{align} 

where $D^{\alpha\beta}$  is the $\alpha\beta$ component of the spin-diffusion tensor. We assume $D^{\alpha\beta} = 0$ for $\alpha \ne \beta$  and  $D^{1,1} = D^{2,2} = D^{3,3} = D$.

\begin{figure}
\centering\includegraphics[width = 3 in]{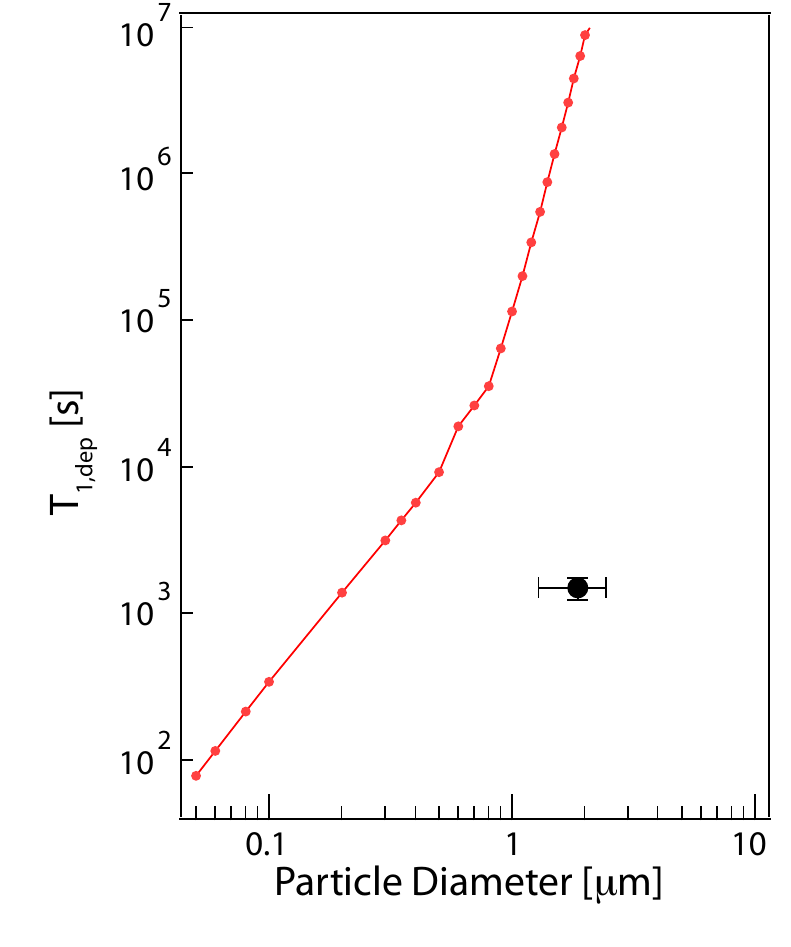}
\caption{\bf{Modelling nuclear depolarization.}\rm~Results of simulations of the nuclear depolarization time constant, $T_{1,dep}$, ~as a function of particle size at 4.7~T and 300~K. The simulations assume a purely crystalline spherical particle with electronic defects located on the particle surface that cause relaxation. The measured room temperature $T_{1}$ of the sample is also shown.
}
\end{figure}

As the depolarizing field is greater than the \siliconspin~nuclear dipolar field (0.08~mT), the nuclear dipolar spin diffusion rate is independent of magnetic field\cite{Abragam}, and for a face centered cubic diamond lattice, the diffusion constant is well described by $D=Wa^{2} \sim a^{2}/(50\,T_2)$, where $a = 0.414$ nm is the average separation between nearest-neighbor nuclei, $W$ is the probability of a flip-flop transition between nuclei due to dipole-dipole interaction, and $T_2 = 10$ ms is the nuclear decoherence time\cite{Khut}. Due to strong gradients in the nuclear Larmor frequency very close to each paramagnetic impurity, nuclear spin diffusion is suppressed ($D = 0$), creating a barrier to diffusion of radius $b = a(\hbar\gamma_{e}^2 B_{\rm dep} / \gamma_{n}2k_{\rm B}T)^{1/4}$\cite{Khut}. Here $\gamma_{n}$ and $\gamma_{e}$ are the nuclear and electronic gyromagnetic ratios, $\hbar$ is the Planck constant and $k_{\rm B}$ the Boltzmann constant.

The contribution from direct relaxation of the nuclear spins by the surface electrons is given by
\begin{align} \left (\frac{dP(r,t)}{dt}\right )_{direct} = \frac{P_0 - P(r,t)}{T_{1n,direct}(r)}, \end{align} 

where 
\begin{align} {T_{1n,direct}(r)}^{-1} = \frac{K}{|R-r|^6} \end{align} 

Here $|R- r |$ is the distance of the nuclear spin packet from the particle surface, and K the strength of the interaction between nuclear and electron spins.

We have previously determined that a model of nuclear relaxation at the particle surface that includes three-spin processes involving pairs of interacting electrons best describes the nuclear spin dynamics in particles similar to these\cite{MYL}. This model gives \begin{align} K= & \frac{3}{10}\frac{\hbar^{2}\gamma_{e}^{2}}{B_{\rm dep}^{2}}  \frac{B_{\rm dep}^{2}\gamma_{n}^{2}T_{2e}}{1+ B_{\rm dep}^{2}\gamma_{n}^{2}T_{2e}^2} \int_{-\infty}^{\infty} \! \frac{g(\omega)g(\omega-\omega_n)}{g(0)} \, \mathrm{d}\omega \end{align} where $T_{2e}$ = 0.025 $\mu$s is the electron spin-spin coupling and $g(\omega)$ is the normalized electron absorption lineshape function\cite {Duijvestijn, Terblanche}. 

This model accounts for flip-flop transitions between nearby electron pairs, occurring on a time scale $T_{2e} \ll T_{1e}$, which provide the fluctuating magnetic field that can flip nuclear spins. For a Lorentzian electron spin resonance lineshape, the integral in (5), which describes the probability of finding two electrons within the ESR line differing in frequency by $\omega_{n}$, can be replaced with the function $2/(4+\omega_{n}^{2}T_{2e}^{2})$.

The average particle size and measured room temperature $T_1$ of the sample used in this study is also plotted in Fig.~S4. The measured value of $T_1$ is significantly lower than what would be expected for a particle with diameter of 2 $\mu$m, however the simulated value of $T_{1,dep}$ agrees well with the 350~nm average crystallite size found by XRD analysis. It should be possible to reduce the actual particle size down to the crystallite size without significantly changing the nuclear spin dynamics reported in this study. Alternatively, the ability to manufacture and polarize large crystalline silicon particles could lead to depolarization times of many hours, or even days.

\section{Magnetic resonance imaging of \siliconspin~nuclei}
\setcounter{subsubsection}{0}
\subsubsection{Imaging methods}
Anatomical imaging was performed with a standard \proton~multiple-slice multiple-echo (MSME) spin echo sequence. The imaging parameters were as follows: coronal orientation, 8 slices, slice thickness = 3 mm, 256 x 256 pixel resolution, field of view (FOV) = 6 cm. A custom fast spin-echo pulse sequence was written for \siliconspin~imaging to take advantage of the particular decoherence properties of solid silicon (see Supplementary Information Fig.~S5). Due to the weak dipolar coupling of the \siliconspin~nuclei in the crystalline lattice, the rapid application of hard $\pi$ pulses in a CPMG sequence can extend the coherence time by several orders of magnitude\cite{Li}. This results in an effective line-width narrowing, which we utilize for imaging the entire 2D image from a single excitation $\alpha$ to the transverse plane.  The imaging parameters were as follows: coronal orientation, single slice, slice thickness = 60 mm, $\alpha = 20^{\circ}-90^{\circ}$, inter-echo time ($\Delta$TE) = 1.48 ms, field of view (FOV) = 40 x 40 mm, 64 x 64 pixel resolution. For images made up of successive averages, the total image repetition time (TR) was 300 ms, which included delays intrinsic to the scanner. The variable flip angle $\alpha$ was chosen based on the number of scans to be performed from a single polarization run so that the magnetization consumed was approximately the same for each excitation.The FOV was increased to 45 x 45 mm for the TRAMP animal images due to the large size of the tumor. 

\begin{figure}
\centering\includegraphics[width = 3 in]{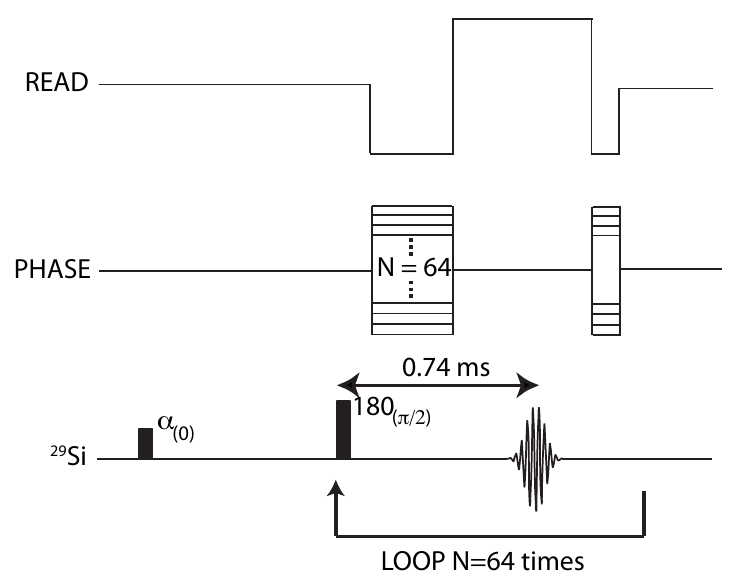}
\caption{Pulse sequence used for \siliconspin~imaging. A single $\alpha$ pulse tipped the magnetization into the transverse plane. Each line of k space was acquired from a spin echo after (N = 64) steps in the phase encode gradient.
}
\end{figure}

\siliconspin~images were exported to Igor Pro (v 6.22, Wavemetrics) for post-processing. Each row of raw k-space data was filtered with a Hanning window to reduce Gibbs ringing and zero-filled to 128 x 128 pixels before a 2D inverse Fourier transform was applied. The magnitude of the resulting image was then taken. Consistent magnitude scaling was used across all phantom and animal experiments. Co-registered images were systematically thresholded so that pixels less than 3 standard deviations above the noise were set transparent and the remaining pixels colored on a gamma function color scale. \proton~images were processed within the Paravision environment.

In Fig.~3(b) the results of 8 successive averages were added together to form the image displayed within the paper. These are all from single polarization events, and used as the intended $\alpha = 90^\circ$ excitation pulse for each image was not accurate across the entire sample and some residual polarization remained. All other images are the result of 1 average.

For the \it{in-vivo}\rm~ images, a 10 dimensional principle component analysis was applied to the magnitude image matrix to reduce the effects of instrumentation and environmental noise affecting the unshielded MRI system\cite{Bankman}.

\siliconspin~phantom images in Fig.~2 were cropped from 40~mm to 15~mm in the horizontal direction after post processing for formatting. This did not remove any visible signal from the image.

\subsubsection{Considerations for \siliconspin~MRI}

The weak dipolar coupling of the \siliconspin~nuclei in the isotopic crystalline lattice results unusual effects not observed in  traditional line-narrowing experiments\cite{Dementyev1, Li, Li2,Dong}. If the pulses are spectrally broad, or strong compared to the internal Zeeman and dipolar spin Hamiltonians, then the delta-function pulse approximation that holds for other materials is no longer true and the internal dipolar spin Hamiltonian can be refocussed. One result of this is a long tail\cite{Li} in the CPMG echo train\cite{Carr, Meiboom}  that is observed for short delays between $\pi$ pulses. This tail persists without decay for several orders of magnitude longer than the $T_2$ as measured with a Hahn echo\cite{Hahn}. 

\begin{figure*}
\centering\includegraphics[width=6 in]{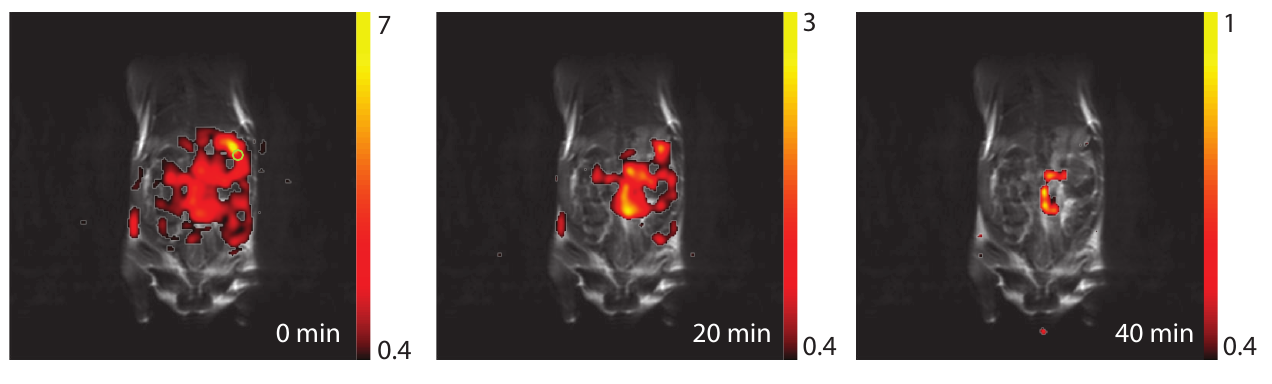}
\caption{\bf{Later stage \siliconspin~MRI intraperitoneal images.}\rm~Later stage \siliconspin~MRI of hyperpolarized silicon particles injected into the peritoneal cavity of the animal (following from Fig. 3B). The green circle shows the approximate location of the catheter in the animal. The particles coat the outside of the stomach and gastrointestinal tract, moving to the center of the animal over longer time periods.}
\vspace{1cm}
\centering\includegraphics[width = 6 in]{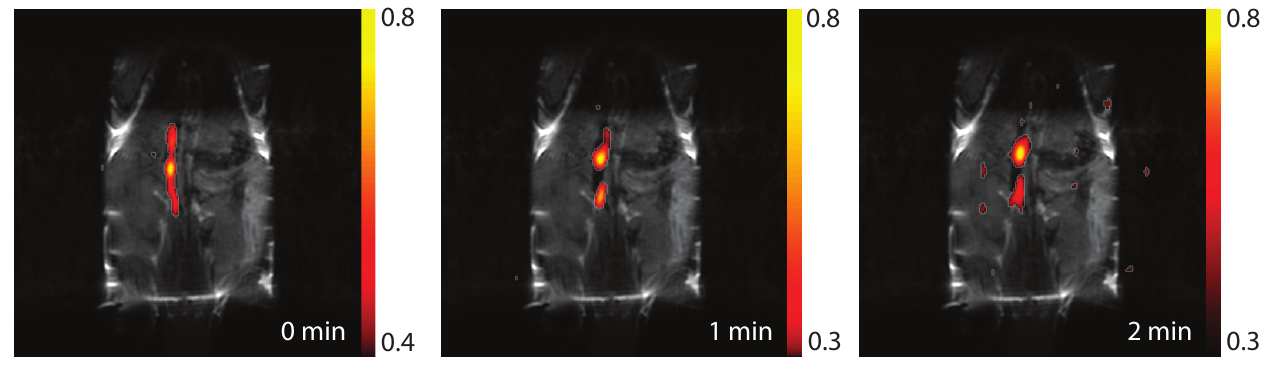}
\caption{\bf{Later stage \siliconspin~MRI intravascular images.}\rm~Later stage \siliconspin~MRI of hyperpolarized silicon particles administered intravascularly via a tail vein catheter (following from Fig. 3C). The vena cava is visible immediately after injection. Images taken two minutes after the injection show particle accumulation in the portal veins and liver and spleen.}
\vspace{1cm}
\centering\includegraphics[width=6 in]{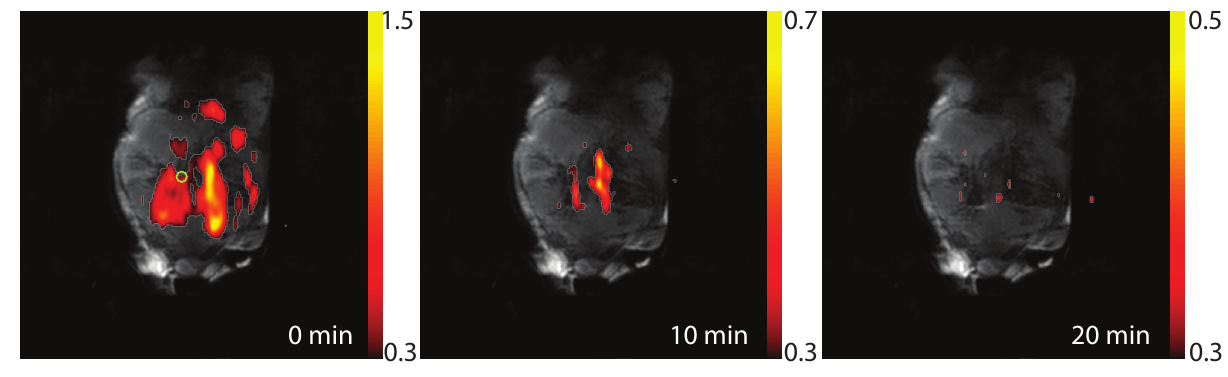}
\caption{\bf{Later stage \siliconspin~MRI perfusion images.}\rm~Later stage \siliconspin~MRI of hyperpolarized silicon particles injected into the tumor of a prostate cancer (TRAMP) animal (following from Fig. 3D). The green circle shows the approximate location of the catheter in the tumor. Initially the silicon particles are more concentrated on the lower right side of the tumor while later images show the silicon particles are more concentrated in the upper right side of the tumor indicating the direction of blood flow from the tumor.}
\end{figure*}

In this work we have used a fast-spin echo (FSE) imaging technique to take advantage of this effect, where a single pulse $\alpha$ excites all the spins in a sample, and then a 2D image is acquired by repeated spin echos together with the application of read gradients and stepping the phase gradient (Fig.~S6). In our implementation only one traverse through k-space is made for each excitation pulse, however, as this 'long-tail' lasts for several hundred milliseconds without decay, k-space averaging could be implemented by repeated cycling of the phase gradient without additional excitation pulses.

A concern for clinical applications of this technology is the specific absorption rate (SAR) issues associated with the use of fast spin-echo pulse sequences. This is certainly true in more aggressive line-narrowing pulse sequences. For example, Dong \it{et al.}\rm~have demonstrated narrowing of NMR linewidth of \siliconspin~by a factor of about 70 000 using quadratic echos\cite{Dong}, and a similar line-narrowing technique has recently been applied in high-resolution ex-vivo MRI of the $^{31}P$ nuclei in bone and brain tissue\cite{Frey}. Fast gradient-echo (FGE) techniques eliminate this issue, however this compromises the signal to noise ratio (SNR) compared to FSE imaging by a factor proportional to the number of lines of k-space addressed and the number of averages. We have however, successfully imaged \siliconspin~nuclei in silicon particles with much lower values of hyperpolarization using these methods\cite{Aptekar}.

One distinct advantage of MRI with a pre-hyperpolarized material such as silicon is that the signal strength is not dependent on the magnetic field, and so imaging could be successfully carried out at much lower magnetic fields where the power deposited by each pulse (which scales quadratically with the nuclear Lamor frequency) is much less. As the \siliconspin~images are overlaid on high-resolution \proton~anatomical scans, the resolution and SNR of the \siliconspin~images can be much lower. This means that less echos have to be applied for a FSE sequence, and less silicon particles must accumulate to provide sufficient signal. 

Unlike regular MRI imaging, our nuclear polarization is high enough so that our in-vivo \siliconspin~images are acquired without averaging, which minimizes the number of echos that have to be applied to a subject within the specified time frame. Up to two orders of magnitude increase in signal could be anticipated by increasing the nuclear polarization, and a factor of 20 increase in signal could be expected from isotopic enrichment.

\section{\textbf{\emph{In-vivo}}~toxicity}
\setcounter{subsubsection}{0}
\subsubsection{\textbf{Toxicity studies}}

Animals (N=2 for each concentration) were injected with concentrations up to the imaged concentrations via tail vein injection, intragastrical feeding needle, or intraperitoneal injection and monitored for a period of two weeks. The concentrations per kg of body weight were 500~\rm mg.kg$^{-1}$, 1500~\rm mg.kg$^{-1}$, 3000~\rm mg.kg$^{-1}$ (IV), 500~\rm mg.kg$^{-1}$, 1000~\rm mg.kg$^{-1}$, 2000~\rm mg.kg$^{-1}$ (IP), 500~\rm mg.kg$^{-1}$, 2500~\rm mg.kg$^{-1}$, 7500~\rm mg.kg$^{-1}$ (GI).
Animals (N=2 for each concentration) were injected with concentrations up to the imaged concentrations via tail vein injection, intragastrical feeding needle, or intraperitoneal injection and monitored for a period of two weeks. The concentrations per kg of body weight were 500~\rm mg.kg$^{-1}$, 1500~\rm mg.kg$^{-1}$, 3000~\rm mg.kg$^{-1}$ (IV), 500~\rm mg.kg$^{-1}$, 1000~\rm mg.kg$^{-1}$, 2000~\rm mg.kg$^{-1}$ (IP), 500~\rm mg.kg$^{-1}$, 2500~\rm mg.kg$^{-1}$, 7500~\rm mg.kg$^{-1}$ (GI).

\subsubsection{Histology} Samples were fixed in 10\% buffered formalin for 24 hours. After fixation the samples were routinely processed and embedded into paraffin blocks. 4 micron sections were then cut and mounted to positively charged slides. 4 step sections at 50 micron intervals were cut. A routine hematoxylin and eosin staining procedure was then performed to visualize the slides. Slides were examined with an Olympus BX51M microscope (Fig.~S8).

\begin{figure}
\centering\includegraphics[width = 3 in]{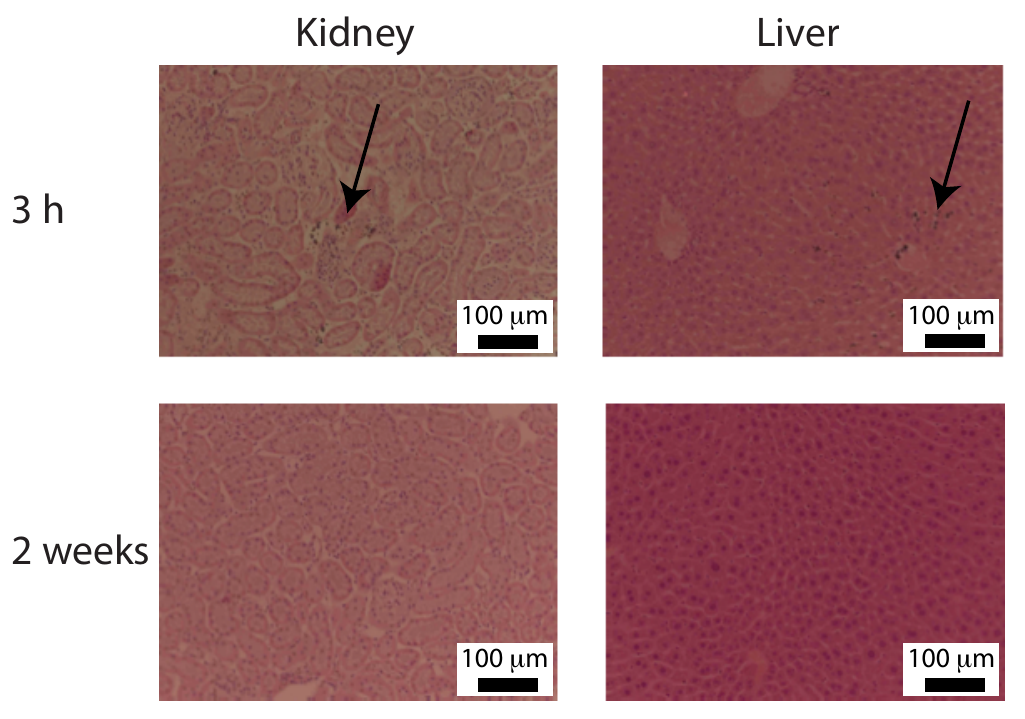}
\caption{\bf{Histology of kidney and liver tissue after intravenous delivery.}\rm~Representative sections of kidney and liver tissue collected from mice three hours and two weeks after intravenous injection of silicon particles. Organs were stained with haematoxylin and eosin. Silicon particles were observable in both the liver and kidney samples collected three hours after delivery (arrows), but not in any of the samples collected two weeks after delivery. The scale bar is 100 $\mu$m for all images.
}
\end{figure}

\end{document}